\documentclass[12pt]{article}
\usepackage{graphicx,amsmath}
\usepackage{units}

\newlength{\dinwidth}
\newlength{\dinmargin}
\def\lapproxeq{\lower .7ex\hbox{$\;\stackrel{\textstyle                                                    
<}{\sim}\;$}}                                                    
\def\gapproxeq{\lower .7ex\hbox{$\;\stackrel{\textstyle                                                    
>}{\sim}\;$}}                                                    
\def\be{\begin{equation}}                                                    
\def\ee{\end{equation}}                                                    
\def\bea{\begin{eqnarray}}                                                    
\def\eea{\end{eqnarray}}

\def\sh{\hat s}
\def\sh2{{\hat s}^2}

\def\CNLO{C^{\rm NLO}}

\def\e{\epsilon}
\def\MS{\overline {\rm MS}}
\def\LL{{\rm LO}\otimes {\rm LO}}
\begin{document}
%\titlepage

\begin{flushright}                                                    
IPPP/13/78 \\
DCPT/13/156 \\                                                    
\today \\                                                    
\end{flushright} 

\vspace*{0.5cm}

\begin{center}
{\Large \bf Physical factorisation scheme for PDFs\\}
\vspace{0.5cm}
{\Large \bf for non-inclusive applications}

\vspace*{1cm}
                                                   
E.G. de Oliveira$^{a,b}$, A.D. Martin$^a$ and M.G. Ryskin$^{a,c}$  \\                                                    
                                                   
\vspace*{0.5cm}                                                    
$^a$ Institute for Particle Physics Phenomenology, University of Durham, Durham, DH1 3LE \\                                                   
$^b$ Instituto de F\'{\i}sica, Universidade de S\~{a}o Paulo, C.P.
66318,05315-970 S\~{a}o Paulo, Brazil \\
$^c$ Petersburg Nuclear Physics Institute, NRC Kurchatov Institute, Gatchina, St.~Petersburg, 188300, Russia \\          
                                                    
\vspace*{1cm}                                                    
                                                    
\begin{abstract}                                                    
We introduce the physical factorisation scheme, which is necessary to describe observables which are {\it not completely inclusive}. We derive the formulae for NLO DGLAP evolution in this scheme, and also for the `rotation' of the conventional $\MS$ PDFs into the physical representation. Unlike, the $\MS$ prescription, where, for example, the gluon PDF at NLO obtains an admixture of the quark-singlet PDF, and vice-versa, the physical approach does not mix parton PDFs of different types.  That is, the physical approach retains the  precise quantum numbers of each PDF. The NLO corrections to DGLAP evolution in the physical scheme are less than those in the $\MS$ case, indicating a better convergence of the perturbative series.
\end{abstract}                                                        
\vspace*{0.5cm}                                                    
                                                    
\end{center}

\section{Introduction}

Pure inclusive observables like deep inelastic scattering, Drell-Yan production, dijet production, are usually described in perturbative QCD via a factorisation theorem, as a convolution of parton distribution functions (PDFs) and coefficient functions calculated in the $\MS$ scheme using dimensional regularisation.  However there are many less inclusive applications in which it is necessary to use PDFs obtained in global analyses of data in a so-called physical scheme.   As examples we mention 
\begin{itemize}
\item exclusive $J/\psi$ and $\Upsilon$ photoproduction, where the cross section is proportional to the square of the gluon PDF, and can provide a unique determination of the gluon at very low $x$ \cite{RY,MNRT,JMRT},
\item central exclusive production of the Higgs boson or dijets or $\gamma\gamma$ or $\chi_c$ etc. \cite{KMRprosp, KMRPLB}
\item the production of a Drell-Yan pair, or prompt large $p_t$ photons, in a limited $p_t$ region; the former process provides a determination of the quark PDF at low $x$ \cite{OMRdy}.
\end{itemize}
These processes are usually described in terms of unintegrated PDFs.

At LO all these observables can be described in terms of conventional $\MS$ partons. However, already at NLO problems appear. The difficulty is that at NLO the $\MS$ scheme mixes partons of different types.
In particular, the gluon distribution contains some admixture of singlet quarks, the charm PDF contains some admixture of gluons and so on. On the other hand, in the physical scheme, where we deal with physical quantities, there is no admixture of parton PDFs of different types. Indeed, when working with Feynman diagrams we have to know the exact quantum numbers corresponding to each propagator, and of each incoming parton. Perhaps the easiest way to see the advantage of the physical scheme is to consider the treatment of the heavy quark distributions \cite{OMRS}. Clearly, for example, the massive charm quark PDF should not obtain an admixture of the gluon PDF in NLO evolution.
Moreover, in the case of NLO Monte Carlo simulations again we have to fix the quantum numbers of each particle\footnote{Recall that in the NLO 
Monte Carlos, where the quantum numbers of each parton must be correctly defined, an alternative scheme to the $\MS$ scheme is used~\cite{MC}.}, and not to consider only inclusive quantities like the transverse energy flow etc. 

Recall that it is necessary to use splitting functions in the physical scheme in order to obtain NLO PDFs unintegrated over the parton transverse momentum \cite{uPDF}, based on the `last step' prescription of \cite{laststep}. 
Of course, it is possible to define unintegrated  PDFs which upon integration lead to the $\MS$ ones. 
 However, this would be a complicated procedure. For each new
process we
would need to calculate a new non-trivial NLO correction to compensate the
non-physical (artificial) contribution introduced by using the $\MS$
scheme, in which we loose the straightforward physical interpretation
given by  Feynman diagrams. For example, it is much easier to account for the heavy quark mass effects by working in the physical scheme, where the mass of each parton is well defined, than by using the $\MS$ scheme, where at NLO (and higher) level we deal with a mixture of different partons. The merits of using the physical scheme for heavy quarks are discussed in detail in \cite{OMRS}. Indeed, for many applications it is very advantageous to have PDFs in the physical scheme.

In Section 2 we explain the origin of the unphysical nature of the $\MS$ PDFs at NLO and higher orders (which is induced by retaining some $\e/\e$ contribution coming fom infinitely large distances). We present the symbolic formula which provides the `rotation' of the known PDFs into the physical PDFs. In Section 3 we present the expressions for the splitting functions which describe the DGLAP evolution in terms of the physical PDFs. The details of the calculations of the NLO splitting functions in the physical scheme are given in the Appendix.

\section{ $ \overline{\bf \rm MS}$ PDFs as ``rotated'' physical partons}

Usually the contributions to NLO DGLAP evolution, and to the corresponding coefficient functions, are calculated in the $\MS$ scheme using dimensional regularisation. Recall that dimensional regularisation is used not only to overcome the ultraviolet divergency of the NLO loop contribution, but also to regularise the infrared (IR) divergency which formally appears in perturbative QCD calculations with massless partons. However, there is a problem with this procedure; some finite $\e/\e$ contributions of IR origin are retained after the subtraction of the $1/\e$ poles. Clearly such contributions are unphysical, since confinement prevents any colour-induced (QCD) interactions at large distances.

In the calculation of the next loop (NLO) diagram, the appropriate procedure is to first subtract the contribution generated by the next step of the LO evolution before performing the integration over $dk^2_t/k^2_t$. Since the IR divergency is of pure logarithmic origin, such a $\LL$ subtraction, which contains exactly the same logarithm, completely cancels the divergency. Thus the remaining NLO contribution may then be calculated in the normal $D=4$ dimensional space.  Here, we call this the `physical' approach. 

Moreover, it was shown in \cite{OMR1} that the NLO coefficient functions, $\CNLO$,  
obtained within the `conventional' $\MS$ prescription using  dimensional ($D=4+2\epsilon$) regularization, are different from the results calculated in the `physical' approach of working in normal $D=4$ space, where, in this case, the infrared divergency is removed  by an appropriate subtraction of the contribution, $C^{\rm LO}\otimes P^{\rm LO}$, generated by the iteration of LO evolution. 

The above difference, $\Delta C$, is due to an $\epsilon/\epsilon$ contribution coming from very large distances. 
It can be written as the convolution 
\be
\Delta C_a \equiv C^{\rm NLO}_a(\MS)-C^{\rm NLO}_a({\rm phys})=\frac{\alpha_s}{2\pi}\sum_b C^{\rm LO}_b\otimes\delta P_{ab}(z),
\ee
 where $a,b=g,q$, and where the $\delta P(z)$ denote the part of the LO splitting functions that are proportional to $\epsilon$ in the $\MS$ approach
\be
P^{\MS}_{ab}(z)=P_{ab}^{\rm LO}(z)+\epsilon \delta P_{ab}(z).
\label{eq:5}
\ee
The $\delta P_{ab}$ are known functions of $z$.  For example, for $q\to q$ splitting we have \cite{CG,OMR1}
\begin{equation}
\delta P_{qq}(x)~=~ C_F \left( (1-x)+ P^{\rm LO}_{qq}(x){\rm ln}(1-x)-\frac{11}{4} \delta(1-x) \right).
\label{eq:3t}
\end{equation} 
The first term comes from the extra gluon polarisation states in $D=4+\epsilon$ space. The second term arises from the phase space factor $(k^2_{t})^\epsilon$, which, when expressed in terms of the virtuality variable, takes the form
\begin{equation}
(k^2_{t})^\epsilon~=~(k^2(1-x))^\epsilon ~=~1+\epsilon{\rm ln}k^2+\epsilon {\rm ln}(1-x).
\label{eq:kte}
\end{equation}
The third term in (\ref{eq:3t}), involving $11/4$, arises from the self-energy diagram. It is necessary to make this term such that it satisfies the conservation of momentum $\int x\delta P (x) dx=0$.

Note that it appears the correction to the NLO coefficient function, $\Delta C$, may be absorbed by redefining the parton distribution\footnote{As seen from (\ref{eq:newparton}), at NLO, the conventional 
 $\overline{\rm MS}$ partons are `rotated' with respect to the physical 
 partons by some angle. In particular, the singlet-quark distribution
gets an admixture of gluons.}
\begin{equation}
a^{\rm phys}(x)~=~a^{\MS }(x)+\frac{\alpha_s}{2\pi}\int\frac{dz}{z}~\delta P_{ab}(z)~b^{\MS}(x/z)~\equiv~a+  \frac{\alpha_s}{2\pi}~\delta P_{ab}\otimes b^{\MS},
\label{eq:newparton}
\end{equation}
where $a=g,q$ ($b=g,q$).
That is, it appears that the $\Delta C$ effect may be considered as adopting an `alternative factorization scheme'. If so, the evolution equation for $a^{\rm phys}(z)$ should follow from $\MS$ evolution.  From (\ref{eq:newparton}) we may write it in the symbolic form
\be
\frac{da^{\rm phys}}{d{\rm ln}Q^2}~=~ \frac{\alpha_s}{2\pi} P \otimes b^{\MS}+ \left(\frac{\alpha_s}{2\pi}\right)^2 \delta P\otimes P \otimes b^{\MS},  
\ee
which, on using (\ref{eq:newparton}) to replace $b^{\MS}$ by $b^{\rm phys}$ gives, neglecting ${\cal O}(\alpha_s^3)$,
\be
\frac{da^{\rm phys}}{d{\rm ln}Q^2}  ~=~  \frac{\alpha_s}{2\pi}P\otimes b^{\rm phys}+ \left(\frac{\alpha_s}{2\pi}\right)^2[\delta P,P]\otimes b^{\rm phys} ~=~ \frac{\alpha_s}{2\pi}\left( P^{\MS}+\frac{\alpha_s}{2\pi}\Delta P\right)  \otimes b^{\rm phys},
\label{eq:13}
\ee
where here $P=P^{\rm LO}+P^{\rm NLO}$. The last equality simply defines $\Delta P$.  Comparing (\ref{eq:13}) with the $\MS$ NLO evolution equation, we see the splitting functions in the physical approach differ from the $\MS$ functions by an amount, $\Delta P$, proportional to the commutator
\begin{equation}
\Delta P \equiv P^{\rm phys}-P^{\MS}=[\delta P,P]~\equiv~\delta P\otimes P-P\otimes \delta P~.
\label{eq:commutator}
\end{equation}
On the other hand, we may calculate the splitting functions directly in the physical approach with $D=4$. Provided we obtain the same $\Delta P$, then the physical approach may be regarded as simply an alternative factorisation scheme\footnote{Unfortunately we made an incorrect statement in a previous paper \cite{OMR2}, where we emphasized the difference between the form of the so-called `+' prescription used in LO evolution and in the calculation of the NLO $\MS$ splitting functions obtained using dimensional regularisation. Indeed, in the second step of the $\LL$ convolution the `+' prescription is written in terms of the ratio $x/x_1$ of the parton's momenta, and not in terms of the cutoff $\delta$ on the energy of the emitted soft gluon, as was done in \cite{CFP} when calculating the NLO diagrams. However, since the `energy' cutoff $\delta$ was implemented consistently in all Feynman diagrams, the soft gluon contribution is cancelled exactly between the real emission and the virtual loop diagrams. Thus, there is no difference between the NLO non-singlet splitting function $P_{qq}$ obtained in the $\MS$ and the `physical' approaches. For the non-singlet case we have $\Delta P \equiv [\delta P_{qq},P_{qq}]=0$. Moreover, it is clear that, due to the exact cancellation of the singular soft gluon contribution, the infrared $\e/\e$ terms do not generate an additional $\Delta P$ between the physical and $\MS$ splittings in singlet evolution as well.
So indeed, the `physical' approach should be regarded as an alternative factorisation scheme, which nevertheless may be very important for many not fully inclusive applications.}, with the two schemes connected by the `rotation' given in (\ref{eq:newparton}).

In Fig. 1 we show the NLO gluon and singlet quark PDFs in the physical scheme, obtained from (\ref{eq:newparton}), by `rotating' the known MSTW partons \cite{MSTW}. Later, we derive expressions for the NLO splitting functions in the physical scheme.
\begin{figure}[t]
\begin{center}
%\vspace*{-3.cm}
\includegraphics[height=7cm]{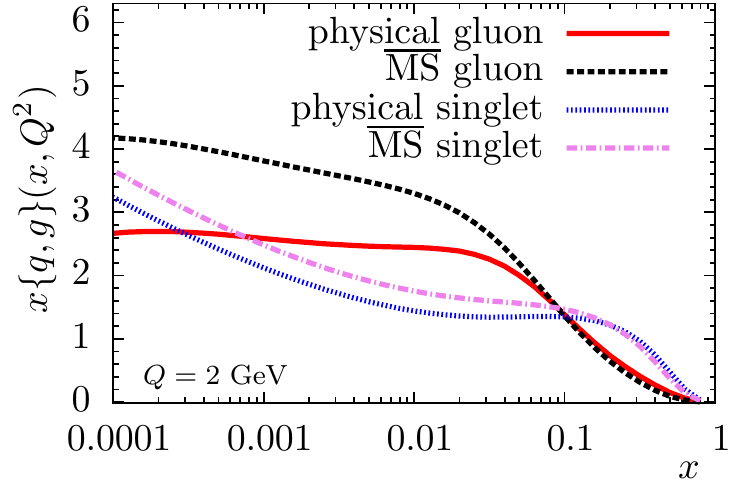}
\includegraphics[height=7cm]{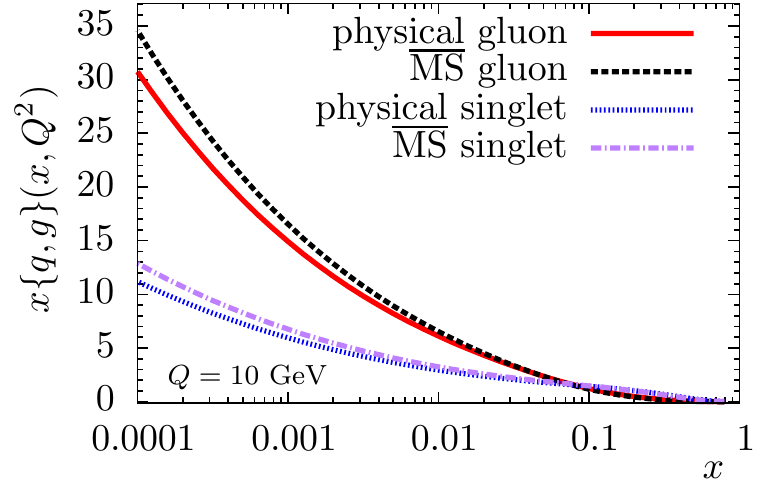}
\vspace*{-0.5cm}
\caption{\sf NLO PDFs in the physical scheme obtained from the MSTW NLO parton set \cite{MSTW} at $Q^2=100~{\rm GeV}^2$} 
\label{fig:1}
\end{center}
\end{figure}

\section{NLO splitting functions in the physical scheme  \label{sec:3}}
Let us return to the terms, $\delta P$, proportional to $\epsilon$ in the LO splitting functions, see (\ref{eq:5}). These functions $\delta P(z)$, can be found, for example, in \cite{CG}. However, in comparison with the results listed in \cite{CG}, we have to add a contribution of pure kinematical origin.
Indeed, in $D=4+2\epsilon$ space the logarithmic integration $\int dk^2_t/k^2_t$ is replaced by $\int d^{2+2\epsilon}k_t/k^2_t\propto (1/\epsilon)(k^2_t)^\epsilon$. If expressed in terms of the virtuality variable, this phase-space factor $(k^2_t)^\epsilon$ takes the form given in (\ref{eq:kte}).
The last term in (\ref{eq:kte}) leads to an additional contribution to $\delta P$ of (\ref{eq:5}) of the form $\e P^{\rm LO}(z)\ln(1-z)$.
Thus we obtain
\be
P^{\rm real}_{qq}(z)=C_F\left[\frac{1+z^2}{1-z}(1+\epsilon\ln(1-z))
+\epsilon(1-z)
%-\epsilon\frac 94\delta(1-z)
\right]\ , 
\label{eq:A1}
\ee
\be
P_{qg}(z)=T_R\left[(z^2+(1-z)^2)(1+\epsilon\ln(1-z))+\epsilon 2z(1-z)\right]\ ,
\ee
\be
P_{gq}(z)=C_F\left[\frac{1+(1-z)^2}z(1+\epsilon\ln(1-z))+\epsilon z\right]\ ,
\ee
\be
P^{\rm real}_{gg}(z)=2C_A\left[\left(\frac z{1-z}+\frac{1-z}z+z(1-z)\right)(1+\epsilon\ln(1-z))
%-\delta(1-z)??? \ . 
\right]\ .
\label{eq:A2}
\ee
Throughout, we use the conventional notation for the colour factors. For QCD this means $C_F=\frac{4}{3},~C_A=3$ and $T_R=\frac{1}{2}$; while $n_F$ denotes the number of light quark flavours.

The complete set of NLO splitting functions, needed to describe the evolution of physical partons, may be obtained from the well-known $\MS$ results by adding on the expressions found in the evaluation of the commutators, $\Delta P \equiv [\delta P,P]$ of (\ref{eq:commutator}), see (\ref{eq:13}).   These expressions for the commutators are listed below, while their detailed evaluation is described in the Appendix.

\subsection{ The commutator for the NLO $q\to q$ and $g\to g$ splitting}

\begin{figure}[t]
\begin{center}
%\vspace*{-3.cm}
\includegraphics[height=5.4cm]{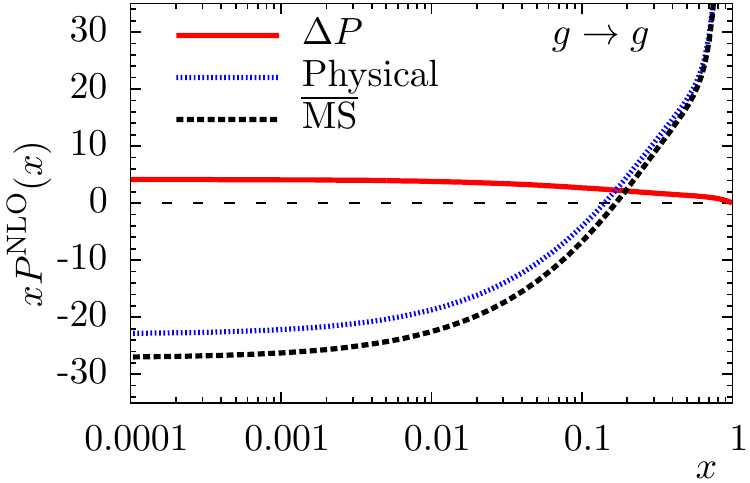} \hfill
\includegraphics[height=5.4cm]{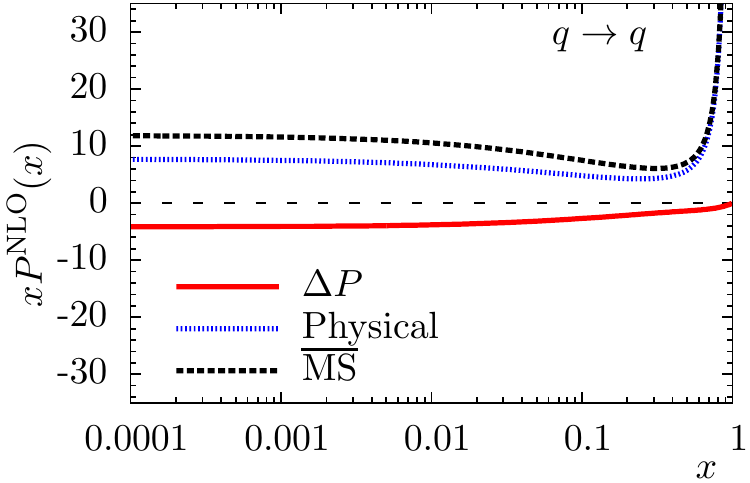}
\includegraphics[height=5.4cm]{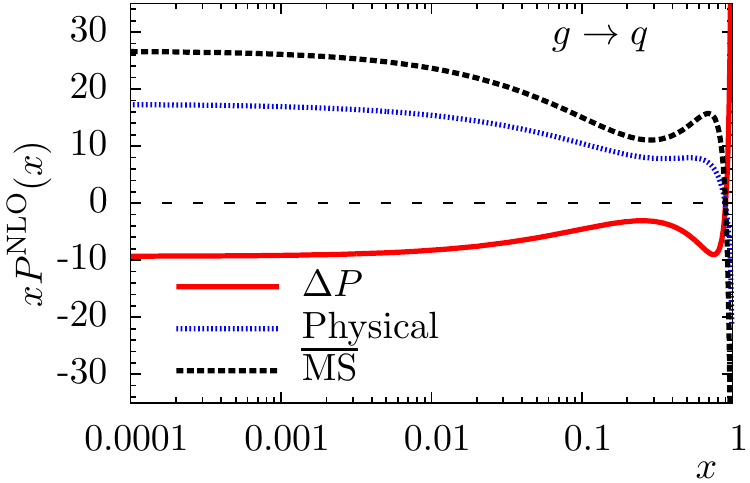} \hfill
\includegraphics[height=5.4cm]{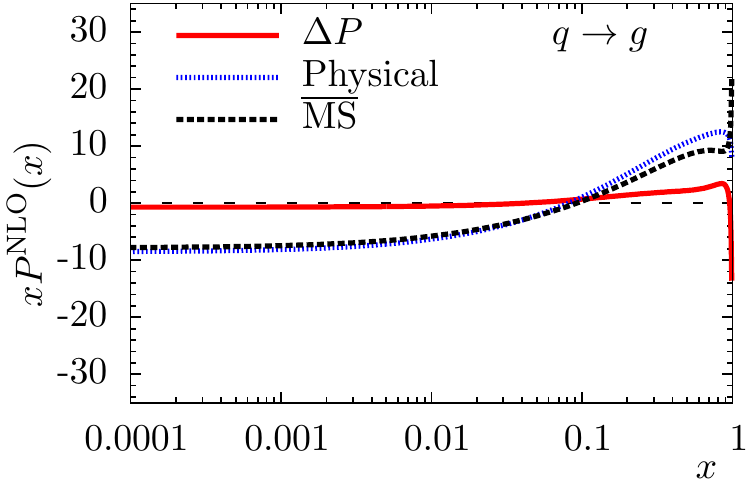}
\vspace*{-0.5cm}
\caption{\sf The NLO splitting functions in physical and $\overline{\mathrm{MS}}$ schemes (dotted--blue and dashed--black lines respectively) and the corresponding difference $\Delta P$ (full--red line).} 
\label{fig:2}
\end{center}
\end{figure}

\begin{align} \nonumber
\Delta P_{qq}\equiv \left[ \delta P, \overline{P}^{\rm LO} \right]_{qq} 
 &~ = ~-\Delta P_{gg}\equiv - \left[ \delta P, \overline{P}^{\rm LO} \right]_{gg} \\
 & =
 2n_F T_RC_F \bigg\{- \frac{ 4 x^2 + 15 x + 6 }{3} \ln (x)
 + \frac{43}{9} x^2 - x - 3
 - \frac{7}{9x} \bigg\},
\end{align}
where $\overline P^{\rm LO}$ indicates that the $\delta(1-x)$ contribution is included in $P^{\rm LO}$.

\subsection{The commutator for the NLO $g\rightarrow q$ splitting}
\begin{align} \nonumber
\Delta P_{qg}\equiv \left[ \delta P, \overline{P}^{\rm LO} \right]_{qg} & = 
 \delta P_{qq} \otimes P^{\rm LO}_{qg} 
 - \overline{P}^{\rm LO}_{qq} \otimes \delta P_{qg} 
  + \delta P_{qg} \otimes \overline{P}^{\rm LO}_{gg} 
 -  P^{\rm LO}_{qg} \otimes \delta P_{gg}
\end{align}
\begin{align} \nonumber
 & = 2 n_F T_R C_F \bigg\{ - (1 +x + x^2)\ln x 
 - \frac{7}{2} x^2 + 6 x - \frac{5}{2}
\\ \nonumber
 & - 4 x (1 - x) \ln (1-x) + 2 P_{qg}(x)
 \left[ - \frac{\ln^2(1-x)}{2}
- \operatorname{Li}_2(1-x) + \frac{\pi^2}{6}
 \right]
\\
 & - \frac{11}{4}  P_{qg}(x)
 - \frac{3}{2} \left[ P_{qg}(x) \ln (1-x)  + 2 x (1-x)\right]
 \bigg\}
\label{eq:part1}
\end{align}
\begin{align}\nonumber
 & + 4 n_F T_R C_A \bigg\{  - \left( \frac{13}{6} x^2 + 4 x + 1 \right)\ln x
 + \frac{257}{36} x^2 - \frac{11}{2} x - \frac{5}{4} - \frac{7}{18 x} \\\nonumber
 & + 2 x (1 - x) \ln (1-x) - P_{qg} (x) 
 \left[ - \frac{\ln^2(1-x)}{2} 
- \operatorname{Li}_2(1-x) + \frac{\pi^2}{6}
 \right] \\
 & - \left[ P_{qg} \ln (1-x) + 2 x(1-x) \right] \left[ \frac{n_F T_R}{3 C_A} - \frac{11}{12}\right]
 + P_{qg}(x) \left[ \frac{203}{144} - \frac{29}{72} \frac{n_F T_R}{C_A} \right] \bigg\}
\label{eq:part2}
\end{align}

\subsection{The commutator for the NLO $q \rightarrow g$ splitting}
\begin{align} \nonumber
\Delta P_{gq} \equiv \left[ \delta P, \overline{P}^{\rm LO} \right]_{gq} & = 
\delta P_{gq} \otimes \overline{P}^{\rm LO}_{qq} 
 -  P^{\rm LO}_{gq} \otimes \delta P_{qq} 
   + \delta P_{gg} \otimes P^{\rm LO}_{gq} 
 -  \overline{P}^{\rm LO}_{gg} \otimes \delta P_{gq}
\end{align}
\begin{align} \nonumber
 & =  C_F^2 \bigg\{ 2 x \ln (1-x) - 2 (2 + x) \ln x  
- \frac{1}{2} x + 5 - \frac{9}{2x}  \\\nonumber
 & - 2 P_{gq} (x)
 \left[ - \frac{\ln^2(1-x)}{2}
- \operatorname{Li}_2(1-x) + \frac{\pi^2}{6}
 \right]
\\
 & +  \frac{3}{2}\left[ P_{gq}(x) \ln (1-x)  + x \right]
 + P_{gq}(x) \frac{11}{4} \bigg\}
\label{eq:part3}
\end{align}
\begin{align} \nonumber
 & + 2 C_A C_F\bigg\{ \left[ \frac{2}{3} x^2 + \frac{5}{2} x+ 2 \right] \ln x
  - \frac{31}{18} x^2 + \frac{5}{4} x - \frac{3}{2} + \frac{71}{36x} \\ \nonumber
 & - x \ln(1-x)  + P_{gq}(x)
 \left[  - \frac{\ln^2(1-x)}{2}
- \operatorname{Li}_2(1-x) + \frac{\pi^2}{6}
 \right]
\\
 & - \left[ \frac{203}{144} - \frac{29}{72} \frac{n_F T_R}{C_A} \right] P_{gq}(x)
  + \left[ \frac{n_F T_R}{3 C_A} - \frac{11}{12}\right]
  \left[ P_{gq} \ln (1-x) + x \right] \bigg\}.
\label{eq:part4}
\end{align}

In Fig. 2 we show the NLO splitting functions in the physical and $\MS$ schemes obtained from the above commutators, $\Delta P \equiv [\delta P,P]$. We also show the difference $\Delta P =P^{\rm phys}-P^{\MS}$.

\section{Conclusions}
We introduce a `physical' factorisation scheme which, unlike the $\MS$ scheme, does not mix partons of different types at NLO.  Note that, already at NLO, the $\MS$ gluons get an admixture of singlet quarks and vice versa. This physical approach should be used to calculate the numerous not fully inclusive processes, such as diffractive $J/\psi$ (or $\Upsilon$) photoproduction, Drell-Yan production in a limited transverse momentum domain, and so on.

The physical scheme also has the advantage that it allows the calculation of unintegrated NLO parton distributions using the `last step' prescription, and that it offers an improved description of heavy quark mass effects during DGLAP evolution \cite{OMRS}.
We give formulae which enable the `physical' PDFs to be obtained by `rotating' the $\MS$ partons known from global analyses. As can be seen from Fig. 1, the difference is not large, but clearly not negligible.

In addition, we derive the formulae giving the difference, $\Delta P \equiv P^{\rm phys}-P^{\MS}$, between the $\MS$ and the `physical' NLO splitting functions. We find that the difference can be as large as 30$\%$, see Fig. 2. Note that, as a rule, $|P^{\rm phys}_{\rm NLO}|<|P^{\MS}_{\rm NLO}|$, which indicates that the perturbative expansion in the physical scheme will have better convergence; that is, the NLO corrections are smaller.

\section*{Acknowledgements}

EGdO and MGR thank the IPPP at Durham University for hospitality. This work was supported by the grant RFBR 11-02-00120-a
and by the Federal Program of the Russian State RSGSS-4801.2012.2;
and by FAPESP (Brazil) under contract 2012/05469-4.

\section*{Appendix}
Here we describe in detail the evaluation of the commutators listed in Section {\ref{sec:3} for singlet evolution. These commutators, $\Delta P_{ab} \equiv [\delta P,P]_{ab}$, give the difference between the NLO splitting functions $P_{ab}$ in the $\MS$ and physical factorisation schemes, see (\ref{eq:commutator}).

%-------------------------------------------------------------------------------
\subsection*{Regularization procedure}
To treat the infrared singularities we follow the procedure used in the original Curci et al paper \cite{CFP}. Now, the intermediate momentum fraction $x_1$ (which plays the role of $z$ in the convolution of (\ref{eq:newparton})) can go from $x$ to 1, and soft divergencies can occur when either $x_1\to 1$ or $x_1\to x$.  Let us start with 
the regularization of the divergence when $x_1\rightarrow 1$, which is treated using a cutoff $\delta$ for the longitudinal momentum carried away by the soft gluon
\begin{align}
\int_x^1 d x_1 \frac{f(x,x_1)}{1-x_1 + \imath \delta}
\end{align}
The real part of above integral is 
\begin{align}
\int_x^1 d x_1 \frac{(1-x_1)}{ (1-x_1)^2 + \delta^2} f(x,x_1),
\end{align}
where in the limit of very small $\delta$, one can use
\begin{align} \nonumber
 \lim_{\delta\rightarrow 0} \int_x^1 d x_1 \frac{(1-x_1)}{ (1-x_1)^2 + \delta^2} f(x,x_1)
  = \int_x^1 d x_1 \frac{f(x,x_1)  - f(x,1)}{1-x_1} 
  + f(x,1) ~[\ln (1-x)+I_0],
\end{align}
so that the singular contribution is collected in a universal function
\be
I_0~=~\int_0^1 du\frac{u}{u^2+\delta^2}=-\ln\delta.
\ee
Similarly when the $x_1\to 1$ divergence is accompanied by a divergent logarithm we may write
\begin{align} \nonumber
\lim_{\delta\rightarrow 0} & \int_x^1 d x_1 \frac{(1-x_1)\ln (1-x_1)}{(1-x_1)^2 + \delta^2}  f(x,x_1) \\
&  = \int_x^1 d x_1 \frac{\ln (1-x_1)}{1-x_1} \left[ f(x,x_1) - f(x,1) \right] 
+ \left[ \frac{\ln^2 (1-x)}{2} + I_1 \right] f(x,1),
\end{align}
where now the singular contribution is collected in another universal function
\be
I_1 ~=~\int_0^1 du\frac{u\ln u}{u^2+\delta^2}= - (\ln^2 \delta)/2 - \pi^2/6 .
\ee

For the case when the divergence happens when $x_1 \rightarrow x$, one has to use the infrared regularisation written in terms of $z=x/x_1$, the variable corresponding to LO evolution. That is
\begin{align}
\lim_{\delta\rightarrow 0} &\int_x^1 \frac{d x_1}{x_1} \frac{f(x,x_1)}{1-x/x_1 + \imath \delta}
 = \int_x^1 d \left( \frac{x}{x_1} \right) \frac{x_1}{x} \frac{f(x,x_1)}{1-x/x_1+\imath\delta} \nonumber \\
 & = \int_x^1 d x_1 \frac{f(x,x_1) - (x/x_1) f(x,x)}{x_1(1-x/x_1)}
 + \left[ \ln (1-x) + I_0 \right] f(x,x).
\end{align}
\begin{align}
\lim_{\delta\rightarrow 0} & \int_x^1 d x_1 \frac{\ln (1 - x/x_1)}{x_1(1-x/x_1 + \imath \delta)} f(x,x_1)
 = \int_x^1 d \left( \frac{x}{x_1} \right) \frac{x_1}{x} \frac{\ln (1 - x/x_1)}{1-x/x_1 + \imath \delta} f(x,x_1)\nonumber \\
 & = \int_x^1 d x_1 \frac{\ln (1 - x/x_1)}{x_1(1-x/x_1)} \left[ f(x,x_1) - \frac{x}{x_1} f(x,x) \right]
 + \left[ \frac{\ln^2 (1-x)}{2} + I_1 \right] f(x,x).
\end{align}
Armed with these results we are able to handle the infrared divergencies and to compute the NLO splitting functions. Note that after accounting for the virtual loop corrections, that is the $\delta(1-z)$ terms in the splitting functions, all the $I_0,~I_1$ contributions cancel, and so there are no divergent terms in the final formulae.

\subsection*{The $\e$ terms, $\delta P$, in the LO splitting function}
Recall from (\ref{eq:5}) that the LO splitting functions in the the $\MS$ scheme have the form
\be
P^{\MS}~=~P^{\rm LO}+\e \delta P,
\ee
see (\ref{eq:A1})$-$(\ref{eq:A2}). A compact form for $\delta P$ corresponding to real emission is
\begin{align}
\delta P^{\rm real} & = \left[ \begin{array}{cc} 
C_F P_{qq}(z) & 2 n_F T_R P_{qg}(z) \\ 
C_F P_{gq}(z) & 2 C_A P_{gg}(z)
\end{array} \right] \ln(1-z)
+ \left[ \begin{array}{cc} 
C_F (1-z)  & 2 n_F T_R 2 z (1-z) \\ C_F z & 0
\end{array} \right].
\end{align}
 We still have to include in the `diagonal' $\delta P$ the virtual contributions which occur at $z=1$. These can be determined from momentum conservation.
For the quark virtual part we obtain
\begin{align} \nonumber
\delta P_{qq}^{\rm virt} & = - \delta(1-z) 
C_F \int_0^1 d z' \, z' \left \{ \left[ P_{qq}(z') + P_{gq}(z') \right] \ln (1-z') + 1 \right\} \\
 & = - \delta(1-z) C_F \left[ \frac{11}{4} + 2 I_1 \right],
\end{align}
while the gluon virtual part is found to be
\begin{align} \nonumber
\frac{\delta P_{gg}^{\rm virt}}{2 C_A} & = - \delta(1-z) 
\int_0^1 d z' \, z' \left \{ \left[P_{gg}(z') + \frac{n_F T_R}{C_A} P_{gq}(z') \right] \ln (1-z') + \frac{n_F T_R}{C_A} 2 z'(1-z') \right\} \\
 & = - \delta(1-z) \left[ \frac{203}{144}  + I_1 - \frac{29}{72} \frac{n_F T_R}{C_A} \right].
\end{align}
Thus we have
\begin{align}
\overline{P}^{\rm LO}_{qq} & = \frac{1+x^2}{1-x + \imath \delta} 
- \delta(1-x) \left[ 2 I_0 - \frac{3}{2} \right] \\
\overline{P}^{\rm LO}_{gg} & = \frac{x}{1-x+ \imath \delta} + \frac{1-x}{x} + x (1-x) 
- \delta(1-x) \left[  - \frac{11}{12} + I_0 + \frac{n_F T_R}{3 C_A}\right].
\end{align}

%-------------------------------------------------------------------------------

\subsection*{Commutators}
Recall that if the commutators $\Delta P \equiv [\delta P,P]$ are added to the well known NLO $\MS$ splitting functions, then we obtain the NLO splitting functions in the physical scheme. We calculate the commutators for the various splitting functions in turn. Below it is implicitly assumed that $P$ denotes the full (real+virtual) splitting function at LO.
%-------------------------------------------------------------------------------
\subsection*{The `diagonal' splittings: $\Delta P_{qq}$ and $\Delta P_{gg}$}
For the $q\to q$ splitting we need to evaluate
\begin{align} \nonumber
\Delta P_{qq} \equiv \left[ \delta P, P \right]_{qq} & = 
\left( \delta P \otimes P -  P \otimes \delta P \right) \big|_{qq} \\ \nonumber
 & = \delta P_{qq} \otimes P_{qq} 
 + \delta P_{qg} P_{gq}
 -  P_{qq} \otimes \delta P_{qq}
 -  P_{qg} \otimes \delta P_{gq}\\
 & = \delta P_{qg} \otimes P_{gq}
 -  P_{qg} \otimes \delta P_{gq},
\end{align}
since $\delta P_{qq} \otimes P_{qq} 
 =  P_{qq} \otimes \delta P_{qq}$. On the other hand for the $g\to g$ splitting we find
\begin{align}
\Delta P_{gg} \equiv \left[ \delta P, P \right]_{gg} 
& = \delta P_{gq} \otimes P_{qg}
 -  P_{gq} \otimes \delta P_{qg}
  = -\Delta P_{qq} \equiv  - \left[ \delta P, P \right]_{qq}.
\end{align}
That is the commutators are identical, except for the sign.

The $gg$ convolutions read
\begin{align}
\left[ \delta P, P \right]_{gg} 
 & = \delta P_{gq}(x/x_1) \otimes P_{qg}(x_1) 
 - P_{gq}(x/x_1) \otimes \delta P_{qg}(x_1) \\ \nonumber
& = C_F 2 n_F T_R\int_x^1 \frac{d x_1}{x_1} \bigg\{ 
 \left[ P_{gq} \left(\frac{x}{x_1} \right)  \ln (1-x/x_1) +  \frac{x}{x_1} \right] P_{qg}(x_1) \\
& - P_{gq}\left(\frac{x}{x_1} \right) \left[ P_{qg}(x_1) \ln (1-x_1) + 2 x_1 (1- x_1)  \right] \bigg\} \\ \nonumber
& = C_F 2 n_F T_R\int_x^1 \frac{d x_1}{x_1} \bigg\{ 
P_{gq}\left(\frac{x}{x_1} \right)  P_{qg}(x_1) \left[ 
 \ln \left( 1 - \frac{x}{x_1} \right) - \ln (1-x_1) \right]  \\
& + \frac{x}{x_1} P_{qg}(x_1) - P_{gq}\left(\frac{x}{x_1} \right) 2 x_1 (1-x_1) \bigg\}.
\end{align}
After evaluating the integrals, the final result is found to be 
\begin{align} \nonumber
\left[ \delta P, P \right]_{qq} 
 & = - \left[ \delta P, P \right]_{gg} \\
 & =C_F 2 n_F T_R \left(
 - \frac{ 4 x^2 + 15 x + 6 }{3} \ln (x)
 + \frac{43}{9} x^2 - x - 3
 - \frac{7}{9x} \right)
\label{eq:f1}
\end{align}

\subsection*{$g\rightarrow q$ splitting, $\Delta P_{qg}$}

Now, no cancellation occurs, and the commutator has more terms
\begin{align} \nonumber
\Delta P_{qg}\equiv \left[ \delta P, P \right]_{qg} & = 
\left( \delta P \otimes P
 - P \otimes \delta P \right) \big|_{qg} \\
 & = \delta P_{qq} \otimes P_{qg} 
 + \delta P_{qg} \otimes P_{gg}
 -  P_{qq} \otimes \delta P_{qg}
 -  P_{qg} \otimes \delta P_{gg}.
\end{align}
It is convenient to  split this sum into two parts corresponding to $g\to q\to q$ and $g\to g\to q$, and to evaluate each separately.  

As an example, we evaluate the first part in detail. That is, we calculate  
 $(\delta P_{qq} \otimes P_{qg} 
 -  P_{qq} \otimes \delta P_{qg})$. These convolutions contain integrals of three different types. First, we evaluate the convolutions containing {\it logarithms}
\begin{align}
\int_x^1 & \frac{d y}{y} \frac{1+y^2}{1-y} 
\left[ \frac{x^2}{y^2} + \left( 1- \frac{x}{y} \right)^2 \right] \left[ \ln (1- y) - \ln \left( 1 - \frac{x}{y} \right) \right] 
\label{eq:39}
\end{align}
where a factor $2 n_F T_R C_F$ is implicit. We start with the part that is most singular $1/(1-y)$ 
\begin{align} \nonumber
 & 2 \left[ x^2 + ( 1- x)^2 \right] \int_x^1 d y \frac{1}{1-y} \left[ \ln (1- y) - \ln \left( 1 - \frac{x}{y} \right) \right] \\
  & = 2 \left[ x^2 + ( 1- x)^2 \right]
 \left[ I_1 - \frac{\ln^2(1-x)}{2} - \ln (1-x) I_0 
- \operatorname{Li}_2(1-x) + \frac{\pi^2}{6}
 \right].
\end{align}
The remaining part of the integrand of (\ref{eq:39}) is
\begin{align} \nonumber
 & \bigg\{\frac{1}{y} \frac{1+y^2}{1-y} \left[ \frac{x^2}{y^2} + \left( 1- \frac{x}{y} \right)^2 \right] - 2 \left[ x^2 + ( 1- x)^2 \right]\frac{1}{1-y}\bigg\} \left[ \ln (1- y) - \ln \left( 1 - \frac{x}{y} \right) \right]\\
  & = \bigg\{-1
  + \frac{4 x^2 - 2 x + 1}{y}
  + 2 x \frac{(x-1)}{y^2}
  +2 \frac{x^2}{y^3}\bigg\}\left[ \ln (1- y) - \ln \left( 1 - \frac{x}{y} \right) \right].
\label{eq:41}
\end{align}
Note that now in (\ref{eq:41}) there is no $1/(1-y)$ pole. Now we use the $(1-y) \leftrightarrow (1-x/y)$ symmetry of the last factor, [...],  to replace $x/y$ by $y$, and then perform the integration. We obtain
\begin{align} \nonumber
 & = \int_x^1 d y \left[
 1 - 2x - 2 y \right]
 \left[ \ln (1- y) - \ln \left( 1 - \frac{x}{y} \right) \right] \\ \nonumber
 & = (1-2x) [ -1 + x - x\ln x] + \left[ x^2 \ln x + \frac{x^2}{2} - 2 x + \frac{3}{2} \right] \\
 & = - (1 - 3 x) x \ln x
   + \frac{1}{2} + x - \frac{3}{2} x^2.
\end{align}

The second component of the convolutions is the {\it non-logarithmic} part.  It can be evaluated to give
\begin{align} \nonumber
\int_x^1 & \frac{d y}{y} \left\{ (1-y) \left[ \frac{x^2}{y^2} + \left( 1- \frac{x}{y} \right)^2 \right] - 
\frac{1+y^2}{1-y} 2 \frac{x}{y} \left( 1 - \frac{x}{y}  \right) \right \}\\
 & = - 4 x (1 - x) [ \ln (1-x) + I_0 ]
 - (1 + 4 x^2)\ln x 
 -2 x^2 + 5 x - 3.
\end{align}
Finally, the {\it third component} containing the $\delta (1-y)$ gives
\begin{align}
 &  - \left[ \frac{11}{4} + 2 I_1 \right] P_{qg}(x)
 + \left[ 2 I_0 - \frac{3}{2} \right] \left[ P_{qg}(x) \ln (1-x)  + 2 x (1-x)\right].
\end{align}

Adding all the components together (now including the factor $2 n_F T_R C_F$) we finally obtain
\begin{align} \nonumber
 & \delta P_{qq} \otimes P_{qg} 
 - P_{qq} \otimes \delta P_{qg} \\ \nonumber
 & = 2 n_F T_R C_F \bigg\{ - (1 +x + x^2)\ln x 
 - \frac{7}{2} x^2 + 6 x - \frac{5}{2}
\\ \nonumber
 & - 4 x (1 - x) \ln (1-x) + 2 P_{qg}(x)
 \left[ - \frac{\ln^2(1-x)}{2}
- \operatorname{Li}_2(1-x) + \frac{\pi^2}{6}
 \right]
\\
 & - \frac{11}{4}  P_{qg}(x)
 - \frac{3}{2} \left[ P_{qg}(x) \ln (1-x)  + 2 x (1-x)\right] \bigg\}.
\label{eq:firstpart}
\end{align}
Note that this only one piece of $\Delta P_{qg}$ given by (\ref{eq:part1}).
The other piece, corresponding to the NLO $g\to g\to q$ splitting, can be evaluated by a similar procedure of dividing the convolutions into three separate components.  It gives the part of $\Delta P_{qg}$ shown in (\ref{eq:part2}).

\subsection*{$q\to g$ splitting, $\Delta P_{gq}$}
A similar procedure may also be used to calculate the $q\to q \to g$ and $q\to g\to g$ pieces of the convolutions arising in $\Delta P_{gq}$.  These contributions are shown explicitly in (\ref{eq:part3}) and (\ref{eq:part4}) respectively.

\subsection*{Momentum conservation}

We may check that our evaluations satisfy momentum conservation. That is, that they satisfy the relations
\be
\int ^1_0 z[\Delta P_{qq}+\Delta P_{gq}] dz=0
\ee
\be
\int ^1_0 z[\Delta P_{gg}+\Delta P_{qg}] dz=0.
\ee
Indeed, we find on integrating our results for the various $\Delta P$'s, that
\begin{align}
\int^1_0 z\Delta P_{qq}dz~&=~-\frac{5}{18} 2n_F C_F T_R,\\
\int^1_0 z\Delta P_{gg}dz~&=~+\frac{5}{18} 2n_F C_F T_R,\\
\int^1_0 z~\Delta P_{qg}({\rm via}~ g\to q\to q)~dz~&=~-\frac{5}{18} 2n_F C_F T_R,\\
\int^1_0 z~\Delta P_{qg}({\rm via}~ g\to g\to q)~dz~&=~~~~~0,\\
\int^1_0 z~\Delta P_{gq}({\rm via}~ q\to q\to g)~dz~&=~~~~~0,\\
\int^1_0 z~\Delta P_{gq}({\rm via}~ q\to g\to g)~dz~&=~+\frac{5}{18} 2n_F C_F T_R,
\end{align}
so that momentum conservation is satisfied for NLO physical evolution, as it must be.

\end{document}